\documentclass[twocolumn]{article}
\usepackage{graphicx,natbib}

\title{\bf Magnetic Fields of Compact Objects\\ in Close X-Ray Binary Systems}
\author{M.Yu. Piotrovich$^1$, Yu.N. Gnedin$^1$, S.D. Buliga$^1$,\\ T.M. Natsvlishvili$^1$, N.A. Silant'ev$^1$, A.S. Nikitenko$^2$\\
(1) Central Astronomical Observatory at Pulkovo of Russian Academy of Sciences, \\Pulkovskoye chaussee 65, Saint-Petersburg, 196140, Russia\\
(2) Saint-Petersburg State Polytechnical University,\\ Politechnicheskaya street 29, Saint-Petersburg, 195251, Russia}
\begin{document}

\maketitle

\begin{abstract}
X-ray binary systems are very popular objects for astrophysical investigations. Compact objects in these systems are neutron stars, white dwarfs and black holes. Neutron stars and white dwarfs can have intrinsic magnetic fields. There is well known, famous theorem about absence of intrinsic magnetic fields of black holes. But magnetic field can exist in the accretion disk around a black hole. We present here the real estimates of the magnetic field strength at the radius of innermost stable orbit in an accretion disk of stellar mass black holes.
\end{abstract}


\section{Introduction}

X-ray binary system consists of normal optical star and a relativistic object. Such relativistic object is a neutron star or a black hole. They are found in the regime of accretion from the optical counterpart. The accretion disk around a relativistic object has high temperature in the central region. It means that the nearest vicinity of the relativistic object is radiating in X-ray range. The problem of the origin of a magnetic field in the innermost parts of the accretion disk remains unclear.

For X-ray binary compact objects the basic regions of magnetic field are accretion disk, relativistic jet and accretion disk wind and outflow. The basic problem is impossibility to use the traditional methods of magnetic field measurement \citep{gnedin00} for accretion disk regions of stellar mass black holes. The main problem is derivation of the magnetic field strength at the innermost stable circular orbit (the ISCO located at the radius $R_{in}$, which presents the radius of the marginal stable orbit).

One of the basic methods for deriving the ISCO magnetic field strength $B_{in}$ is to use the kinetic power of a relativistic jet. Namely the kinetic power of a relativistic jet depends strongly on the magnetic field strength at ISCO and on the spin of black hole.

There exists a number of methods for determination of kinetic power of relativistic jet. For example, this power can be determined from data about high-energy gamma ray luminosity presented by Fermi-LAT cosmic observatory.

\section{Observational constraints on the mechanism of generation of relativistic jets in X-ray binary systems}

In X-ray binaries compact jets are known to commonly radiate at frequencies form radio to infrared \citep{russel13,russel14}. In the radio to optical bands, a strong non-thermal component was associated with synchrotron emission from a powerful jet. Traditionally it is considered that the spectral energy distribution from optical to EUV bands is dominated by a thermal radiation from the accretion disk. The X-ray emission consists of a typical power-law spectrum with photon index $\Gamma \sim 1.8$ \citep{malzac04}. The X-ray properties of the source and the presence of strong radio emission are really typical for black hole binaries in the hard state. Certainly, the jet power is dependent essentially on black hole spin. The detailed review of jets and outflows in stellar mass black holes have been recently presented by \citet{fender14}. For our estimations of the ISCO magnetic field we used the results of dependence of BH spin on relative jet power from \citet{russel13}. Presented at Fig.2 of \citet{russel13} the value $P_j$ is the ratio of the real kinetic power value $L_j$ to the Eddington luminosity of a black hole, i.e. $P_j = L_j / L_{Edd}$. The jets, that are observed in the black hole X-ray binaries (BHXBs), usually manifest themselves as powerful compact outflows commonly seen as a flat or slightly inverted radio spectrum. In BHXBs most of energy radiated by these jets is emitted in infrared-optical regime and even higher X-ray and $\gamma$-ray energies in some cases.

\citet{corbel11} have demonstrated existence of radio and X-ray flux correlation in the hard and quiescent states (see Fig.9 from their paper). Their new measurement represent the largest sample for a stellar mass black hole, without any bias from distance uncertainties, making GX 339-4 the reference source for comparison with other accreting sources. These results demonstrate very strong and stable coupling between radio and X-ray emission. The radio and X-ray luminosity correlation of the form $L_x \sim L_R^{0.62\pm 0.01}$ confirms the non-linear coupling between the jet and the inner accretion flow power.

There is the empirical evidence for a connection between accretion and outflow in X-ray binaries \citep{fender14}. Thus the relation exists also between jets and X-ray states and clear relation exists between accretion disk winds and the soft X-ray states.

Recently, a positive correlation between spin parameter and jet power for transient jets was confirmed by \citet{narayan12} and \citet{steiner13}.

The ''spin paradigm'', whereby high black hole spin can lead to powerful jet production, have been carefully investigated by \citet{garofalo10}. Under the assumption that jets are produced by the combined effort of the Blandford-Znajek (BZ) and Blandford-Payne (BP) \citep{blandford77,blandford82} effects, recent numerical studies of general relativistic magnetohydrodynamics of black hole accretion flows suggest a tight link between jet power and black hole spin \citep{garofalo09,garofalo10}. Whereas BZ process involves jets produced via spin-energy extraction from the rotating black hole, BP process is a jet mechanism that originates in accretion disk via mass-loading of disk plasma onto large scale of magnetic field that is concentrated near the innermost stable circular orbit $R_{in}$. The black hole spin dependencies of these processes have direct implications for stellar mass black holes in X-ray binary systems.

For combined effect of BZ and BP processes the kinetic power of relativistic jet can be presented in a form \citep{garofalo10}:

\begin{equation}
 L_j = 10^{-16} B_{in,5}^2 \left(\frac{M_{BH}}{10 M_{\odot}}\right)^2 f(a),
 \label{eq1}
\end{equation}

\noindent where $L_j$ is the kinetic power of a relativistic jet, $M_{BH}$ is the black hole mass, $B_{in,5} = B_{in} / 10^5\,G$, $f(a)$ is the function depending on the dimensionless spin $a$ of a black hole. This function is tabulated in \citet{garofalo10} (Fig.3 from this paper). \citet{garofalo10} presented also the analytical relation for the function $f(a)$:

\begin{equation}
 f(a) = 2 \times 10^{47} \alpha \beta^2 a^2,
 \label{eq2}
\end{equation}

\noindent where $\alpha = \delta (3/2 - a)$ and

\[
 \beta = - \frac{3}{2} a^3 + 12 a^2 - 10 a + 7 - \frac{0.002}{(a - 0.65)^2} +
\]
\begin{equation}
 + \frac{0.1}{a + 0.95} + \frac{0.002}{(a - 0.055)^2}.
 \label{eq3}
\end{equation}

\noindent The $a$ spans negative values for retrograde spin and positive values for prograde spin, and besides a conservative value $\delta = 2.5$ have been adopted. Using then the real data of relation between jet power and spin value from \citet{russel13} and \citet{russel14} it is possible to estimate the value of the ISCO magnetic field strength $B_{in}$ for the stellar mass black holes in X-ray binaries.

\section{The ISCO magnetic fields of black holes in Cyg X-1 and GRS 1915+105}

The basic parameters of the compact object in the X-ray binary Cyg X-1 are the black hole mass $M_{BH} = (14.8\pm 1.0)M_{\odot}$, the dimensionless spin $a = 0.96\pm 0.04$ \citep{cherepashchuk14} and the relativistic jet kinetic power $L_j = 10^{-2.5} L_{Edd}$, where $L_{Edd}$ is the Eddington luminosity. Using the Eq.(\ref{eq1}), we obtain the next estimate for the ISCO magnetic field strength $B_{in} = 10^{6.82} G \approx 10^7 G$.

\citet{karitskaya10} have been able to measure the magnetic field in the region of generation HeII line $\lambda 4868$ emission. They obtain $B(R_{\lambda}) = 449\, G$, where $R_{\lambda}$ is the radius of the accretion disk where HeII line is generating. For the standard \citep{shakura73} accretion disk the distance $R_{\lambda}$ is derived by relation \citep{poindexter08}:

\begin{equation}
 R_{\lambda} = 0.97 \times 10^{10} \lambda^{4/3} (\mu m) \left(\frac{M_{BH}}{M_{\odot}}\right)^{2/3} \left(\frac{l_E}{\varepsilon}\right)^{1/3},
 \label{eq3_1}
\end{equation}

\noindent where $\lambda$ is the wavelength of the accretion disk emission, $\varepsilon$ is the radiation emissivity of the accretion disk, $l_E = L_{bol} / L_{Edd}$, $L_{bol}$ is the bolometric luminosity and $L_{Edd}$ is the Eddington luminosity. Suggesting the power-law dependence of the magnetic field in the accretion disk in Cyg X-1, i.e. $B(R_{\lambda}) = B_{in}(R_{in} / R_{\lambda})^n$, one can estimate the power-law index $n$. For the spin value $a = 0.96$ the ISCO radius is $R_{in} = 10^{6.54}\, cm$ and the coefficient of radiation emission is $\varepsilon (a) = 0.18$ \citep{hawley06}. For measured magnitude of magnetic field in the region of $\lambda 4686$ emission $B(R_{\lambda}) = 449\, G$ the index $n = 1.4 \approx 4/3$. It is interesting that this value is inversely proportional to corresponding index of the effective temperature dependence for the standard accretion disk: $T_e \sim R^{-3/4}$.

For GRS~1915+105 we have $M_{BH} = 14 M_{\odot}$, $a = 0.975$, $L_j = 10^{-0.25} L_{Edd}$ and $B_{in} = 10^{7.96}\, G \approx 10^8\, G$. we also estimated the ISCO magnetic field by other way, using the results of \citet{piotrovich11}. In this paper the empirical relation between the characteristic frequency of quasi-periodic oscillator (QPO) $\nu_{br}$ has been used. There is a close relation between QPO frequency $\nu_{br}$ and the magnetic field strength at the horizon radius $B_H$ \citep{piotrovich11}:

\begin{equation}
 B_H = \frac{\sqrt{k \nu_{br}}}{1 + \sqrt{1 - a^2}},
 \label{eq4}
\end{equation}

\noindent where $k = P_{mag} / P_{acc}$ is the ratio between magnetic and accretion gas pressures at the horizon radius.

According to \citet{garofalo09} the central inward plunging region of the accretion flow enhances the trapping of large scale poloidal field on the event horizon of the black hole. As a result the ratio of horizon threading magnetic field $B_H$ as measured by ZAMO observers and the magnetic field strength in the accretion disk $B_{in}$ is the function of spin $a$ \citep{garofalo09}. This result corresponds to the next relation:

\begin{equation}
 B_H = \chi(a) B_{in} \approx e^{2a} B_{in}.
 \label{eq5}
\end{equation}

\noindent The characteristic QPO frequency of GRS~1915+105 is $\nu_{br} = 168\, Hz$, and according to (\ref{eq4}) we obtain $B_{in} = 4.8 \times 10^8\, G$ in a situation when $k = 1$ and $a = 0.99$. Then Eq.(\ref{eq5}) gives the next value for the ISCO magnetic field $B_{in} = 10^{7.84}\, G$, that is close to the value $B_{in} = 10^{7.96}\, G$ obtained from Eq.(\ref{eq1}) with the jet kinetic power $L_j = 10^{-0.25} L_{Edd}$ \citep{russel13}.

\sloppy Results of our calculations of the ISCO magnetic field strength $B_{in}$ for other compact objects in X-ray binaries are presented in the Table 1.

\fussy

\begin{table}[htpb]
\footnotesize
\caption{Results of our calculations of the ISCO magnetic field strength for compact objects in X-ray binaries.}
\begin{center}
\begin{tabular}{lccc}
\hline
\textbf{Object}&
{\bf M} [$M_\odot $]&
\textbf{Spin}&
{\bf B} [G] \\
\hline
Cyg X-1&
14.81$\pm $0.98&
0.96$\pm $0.04&
$10^{6.82}$ \\
Gro 1655-40&
6.3$\pm $0.3&
0.7$\pm $0.1&
4.47$\times 10^8$ \\
GRS 1915+105&
14$\pm $4&
0.975$\pm $0.025&
$10^{7.96}$ \\
XTE J1550-564&
9.6$\pm $1.2&
0.55$_{-0.22}^{+0.15} $&
$10^{8.326}$ \\
XTE J1652-453&
10&
0.5&
$10^{7.805}$ \\
A0620-00&
6.6$\pm $0.25&
0.12$\pm $0.19&
$10^{8.87}$ \\
GX 339-4&
12.3$\pm $1.4&
0.94&
$10^{8.26}$ \\
H 1743-322&
13.3$\pm $3.2&
0.2$\pm $0.3&
$10^{8.895}$ \\
XTE J1752-223&
9.5$\pm $0.9&
0.52$_{-0.13}^{+0.16} $&
$10^{8.565}$ \\
GS 2000+25&
7.5$\pm $0.3&
$\sim 0.03$&
$10^{7.64}$ \\
GRS 1124-68&
7.3$\pm $0.8&
$\sim -0.04$&
$10^{8.865}$ \\
{\scriptsize MAXI J1836-194}&
10.0&
0.88$\pm $0.03&
$10^{6.81}$ \\
{\scriptsize XTE J1908+094}&
10.0&
0.75$\pm $0.09&
$10^{7.21}$ \\
433 X-7&
15.55$\pm $3.20&
0.84$\pm $0.05&
{\scriptsize $10^{8.41}\left(\frac{L_j}{L_{Edd}}\right)^{\frac{1}{2}}$} \\
4U 1957+11&
16.0&
$\sim 0.9$&
{\scriptsize $1.6\times 10^8\left(\frac{L_j}{L_{Edd}}\right)^{\frac{1}{2}}$} \\
\hline
\end{tabular}
\label{tab1}
\end{center}
\end{table}

\section{Conclusions}

 As a result we presented the real estimates of the magnetic field strength at the radius of innermost stable circular orbit in an accretion disk of stellar mass black holes. The basic method for deriving the ISCO magnetic field strength $B_{in}$ is to use the kinetic power of a relativistic jet in X-ray binary systems. For estimation of the kinetic power we have used the results of the paper by \citet{russel13}. In BHXBs most of energy radiated by these jets is emitted in infrared, X-ray, and $\gamma$-ray energies. For our estimations we used also the positive correlation between spin parameter and jet power. It appears that the high black hole spin value can lead to the production of powerful jets. We used also the assumption that jets are produced by the combined effect of the Blandford-Znajek and Blandford-Payne processes of relativistic jet generation. Obtained values of the ISCO magnetic fields for the stellar mass black holes are presented in Table 1. On can see that the typical values of ISCO magnetic field for the stellar mass black holes are in the limits $B_{in} \sim 10^7 \div 10^8\, G$. It is interesting to compare these values with corresponding values for supermassive black holes. According to \citet{silantev09,silantev11,silantev13} the typical values of the ISCO magnetic field for SMBHs are $B_{in} \sim 10^4 \div 10^5\, G$. It means that there exists the physical mechanism of generation of magnetic field in accretion disks around black holes which strongly depends on the black hole mass.

\section*{Acknowledgments}

This work was supported by the Basic Research Program of the Presidium of the Russian Academy of Sciences P-21 ''Non-stationary phenomena in objects of the Universe'' and the Basic Research Program of the Division of Physical Sciences of the Russian Academy of Sciences OFN-17 ''Active processes in Galactic and extragalactic objects''.

\end{document}